\documentclass[prl,twocolumn,superscriptaddress,reprint, amsmath,amssymb]{revtex4-1}

\usepackage{graphicx}
\usepackage{dcolumn}
\usepackage{bm}

\newcommand{\beq}{\begin{equation}}
\newcommand{\eeq}{\end{equation}}

\renewcommand{\rho}{\varrho}
\renewcommand{\theta}{\vartheta}
\renewcommand{\phi}{\varphi}

\newcommand{\wegdamit}[1]{} 

\newlength{\lwveryfine}   
\newlength{\lwfine}   
\newlength{\lwnormal} 
\newlength{\lwthick}  
\newlength{\lwverythick}  

\begin{document}

\preprint{APS/123-QED}

\title{Low Energy Spread Attosecond Bunching and Coherent Electron Acceleration in Dielectric Nanostructures}

\author{Uwe Niedermayer}
\email{niedermayer@temf.tu-darmstadt.de}
\affiliation{%
Technische Universit\"at Darmstadt, Institut f\"ur Teilchenbeschleunigung und Elektromagnetische Felder (TEMF), Schlossgartenstrasse 8, D-64289 Darmstadt, Germany
}%

\author{Dylan S. Black}
\email{Uwe Niedermayer and Dylan S. Black equaly contributed to this work.}

\author{Kenneth J. Leedle}
\author{Yu Miao}
 \affiliation{Department of Electrical Engineering, Stanford University, 350 Serra Mall, Stanford, California 94305-9505, USA}
\author{Robert L. Byer}
\affiliation{Department of Applied Physics, Stanford University, 348 Via Pueblo Mall, Stanford, California 94305-4090, USA}
\author{Olav Solgaard}
 \affiliation{Department of Electrical Engineering, Stanford University, 350 Serra Mall, Stanford, California 94305-9505, USA}

\date{\today}

\begin{abstract}
We demonstrate a compact technique to compress electron pulses to attosecond length, while keeping the energy spread reasonably small. The technique is based on Dielectric Laser Acceleration (DLA) in nanophotonic silicon structures. Unlike previous ballistic optical microbunching demonstrations, we use a modulator-demodulator scheme to compress phase space in the time and energy coordinates. With a second stage, we show that these pulses can be coherently accelerated, producing a net energy gain of $1.5\pm0.1$~keV, which is significantly larger than the remaining energy spread of $0.88 \,_{-0.2}^{+0.0}$~keV FWHM. We show that by linearly sweeping the phase between the two stages, the energy spectrum can be coherently moved in a periodic manner, while keeping the energy spread roughly constant. After leaving the buncher, the electron pulse is also transversely focused, and can be matched into a following accelerator lattice. Thus, this setup is the prototype injector into a scalable DLA based on Alternating Phase Focusing (APF).

\end{abstract}

\pacs{Valid PACS appear here}
\maketitle


Dielectric Laser Acceleration (DLA) provides the highest gradients among structure based particle accelerators by utilizing the GV/m femtosecond laser damage thresholds of nanostructured dielectric materials. After the first proposals~\cite{Shimoda1962ProposalMaser,Lohmann1962ElectronWaves}, it took 50 years for the first experimental DLA demonstrations to be realized ~\cite{Peralta2013DemonstrationMicrostructure.,Breuer2013DielectricEffect}. Recently, the gradients have been further pushed to 690~MeV/m~\cite{Wootton2016DemonstrationPulses} and 850~MeV/m~\cite{Cesar2018High-fieldAccelerator} for relativistic electrons and to 133~MeV/m~\cite{Yousefi2019DielectricReflector} and 370~MeV/m~\cite{Leedle2015LaserStructure} for their low-energy counterparts.
While the relativistic experiments use an RF photoinjector and sometimes multiple RF-preaccelerator stages, the subrelativistic experiments require an ultra-low emittance nanotip-emitter~\cite{Ehberger2015HighlyTip,Feist2017UltrafastBeam,Tafel2019FemtosecondTips,AndrewCeballos2019SiliconAccelerators, Hirano2020AAccelerator} and an electrostatic preaccelerator to obtain suitable electron beams for injection into DLAs. Especially at these low injection energies, a beam confinement and bunching scheme~\cite{Naranjo2012StableHarmonics,Niedermayer2018Alternating-PhaseAcceleration,Niedermayer2020ThreedimensionalAccelerators} is required to scale DLAs up to MeV energy gain for various applications~\cite{Zewail20104DMicroscopy,Egerton2015OutrunElectrons,Morimoto2018DiffractionTrains}.

With unbunched electron beam injection, the DLA energy spectra show the typical symmetric shoulder modulation~\cite{Peralta2013DemonstrationMicrostructure.,Wootton2016DemonstrationPulses, Yousefi2019DielectricReflector, Sapra2020On-chipAccelerator}, which can be analytically modeled using probability theory~\cite{Egenolf2019AnalyticalAccelerators}. 
Combined with on-chip ballistic bunching, net-acceleration and steering in a downstream DLA stage was recently demonstrated~\cite{Black2019NetAccelerator,Schonenberger2019GenerationAcceleration}. However, the large energy spread created in the buncher stage quickly dissolves the microbunch phase coherence. In this Letter, we demonstrate an optical modulator-demodulator that compresses first the bunch length and then the energy spread. The resulting simultaneously ultra-short and low energy spread electron pulses can be captured in a potential well ("bucket") and accelerated in a lossless and scalable fashion. 

At low injection energies from typical nanotip electron sources ~\cite{Ehberger2015HighlyTip,Feist2017UltrafastBeam,Hirano2020AAccelerator}, the Alternating Phase Focusing (APF) scheme~\cite{Niedermayer2018Alternating-PhaseAcceleration} is well suited to confine and accelerate the beam, since it is flexible in focusing cell design and economic in field strength to acceleration gradient conversion, which constitutes the performance bottleneck of a scalable accelerator~\cite{Niedermayer2020ThreedimensionalAccelerators}. 
However, as compared to constant longitudinal focusing~\cite{Niedermayer2017DesigningChip}, which does not provide any transverse confinement, the temporal acceptance of APF is slightly smaller. The structure we present here is an APF-based buncher, which is suitable to inject into an APF-based accelerator, due to matching of both the sub-fs bunch length and the low energy spread.


Attosecond pulses of subrelativistic electron beams can be created by ballistic bunching~\cite{Niedermayer2017DesigningChip,Morimoto2018DiffractionTrains,Black2019NetAccelerator,Schonenberger2019GenerationAcceleration}, i.e., a sinusoidal energy modulation  $\Delta W$ is turned to bunching after reaching the longitudinal focal length of 
\begin{equation}
    L=\frac{\lambda_g \beta^2\gamma^3 m_e c^2}{2\pi \Delta W},
\end{equation}
where $\lambda_g$ is the period length of the modulation, $\beta$ is the injection velocity in units of $c$, $\gamma$ is the mass factor, and $m_e c^2$ is the electron rest energy. Note that if the modulation is produced by a laser of wavelength $\lambda$ in a DLA grating, the Wideroe resonance condition $\lambda_g=\beta\lambda$ has to be fulfilled, where $\lambda_g$ is the grating period.

The motion towards the focus is linear only in the vicinity of the fixed point(s) and strongly nonlinear elsewhere. Thus, the longer the dispersive drift after the modulation, the more irreversible longitudinal emittance growth is produced. Moreover, since the modulation is essentially longitudinal focusing, according to Earnshaw's theorem~\cite{Earnshaw1842OnEther} transverse defocusing comes along with it. The idea of the setup presented here is to introduce a grating segment which removes these drawbacks by removing the modulation after a certain length of drift, which is shorter than the focal length. The final longitudinal focus is then reached with a significantly smaller energy spread and transversely focused.

\begin{figure}[t]
\centering
\includegraphics[width=0.49\textwidth]{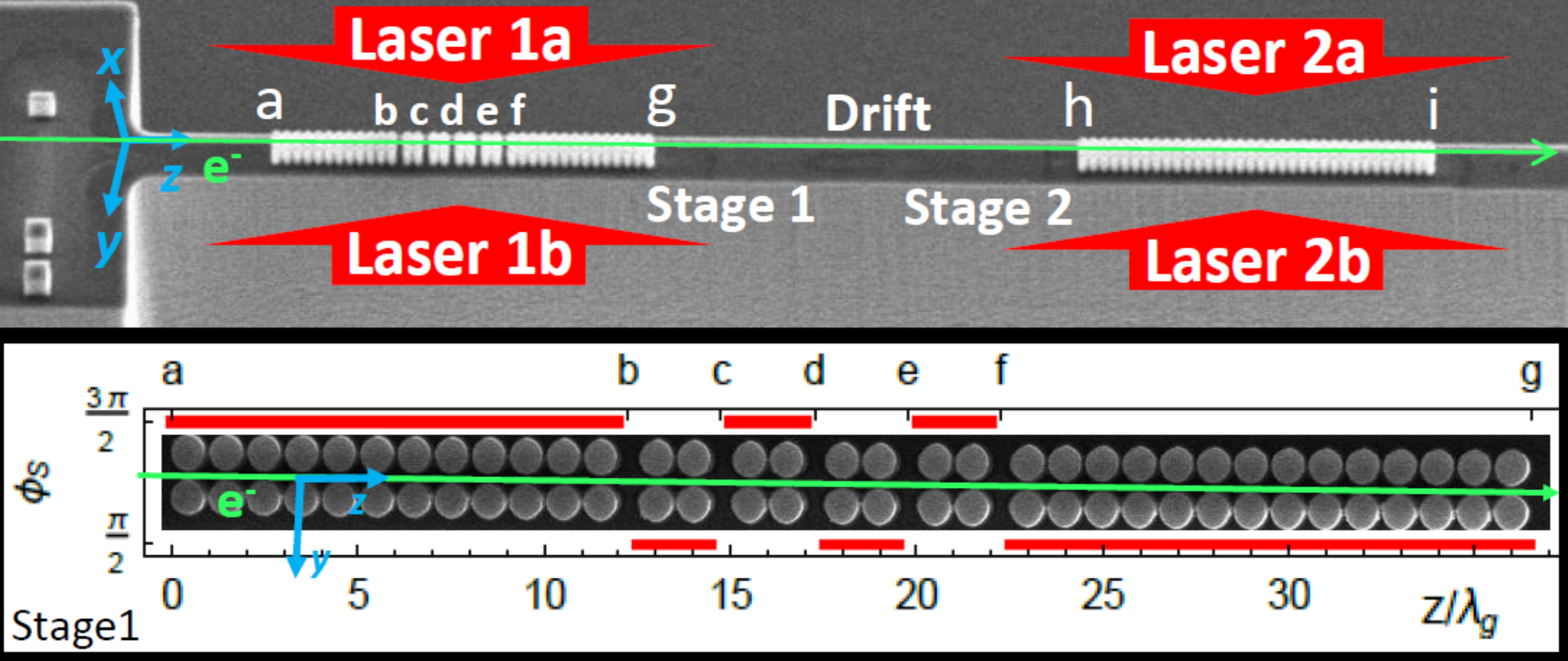}
\caption{Scanning electron microscope image of the structure. The inset shows an enlarged top-view of the buncher stage with the designed distribution of the synchronous phase.}
\label{Picture}
\end{figure}

The phase acceptance of a scalable APF-DLA depends on the choice of the synchronous phase of the accelerator, i.e., in an overall design one has to trade off temporal acceptance and resulting average gradient. At $\pm60^\circ$ off-crest, where the gradient is half the peak gradient, the full temporal acceptance in a longitudinal focus is about 5\% of an optical cycle~\cite{Niedermayer2018Alternating-PhaseAcceleration}, which is $\delta t=330$~as for the 1980~nm driver laser we use here. The matched energy spread in an APF lattice is (see~\cite{Niedermayer2018Alternating-PhaseAcceleration} supplemental Eq.~18)
\begin{equation}
    \delta W=m_ec^2 \beta^3\gamma^3\frac{c}{ \hat\beta_L} \delta t,
    \label{Eq:Mathcing}
\end{equation}
where $\hat\beta_L$ is the longitudinal Courant-Snyder beta-function at the beginning of the accelerator. The minima of $\hat\beta_L$ can reach down to $10-20~\mu$m, depending on the laser amplitude on which the APF accelerator is driven. At a reference energy of 57~keV, the resulting full energy spread acceptance is 286-572~eV and can be filled by an injector as presented here.

\begin{figure}[h!]
\centering
\includegraphics[trim = 5mm 4mm 5mm 4mm, clip, 
width=0.5\textwidth]{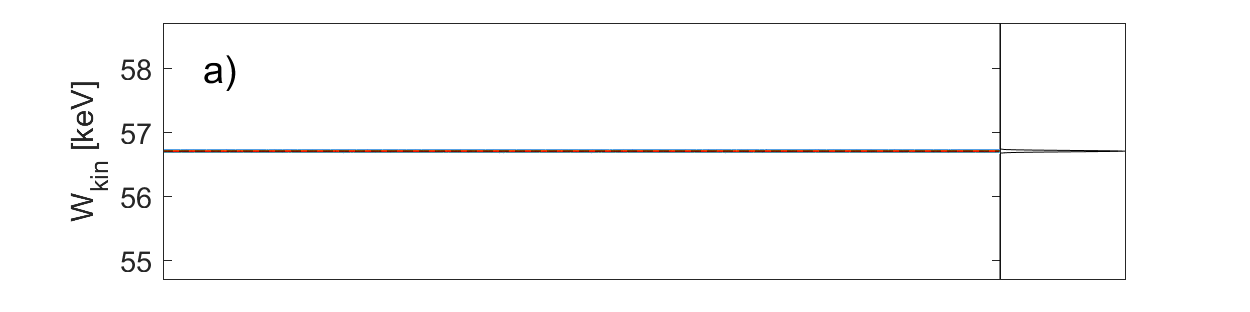}
\includegraphics[trim = 5mm 4mm 5mm 4mm, clip, 
width=0.5\textwidth]{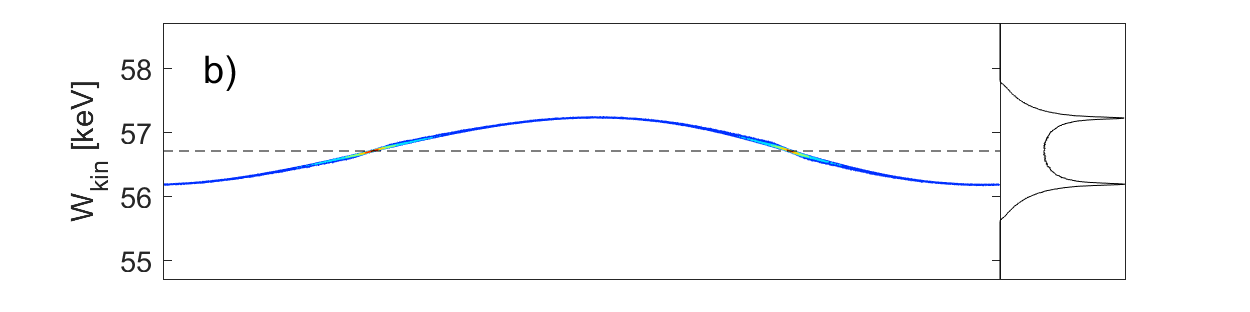}
\includegraphics[trim = 5mm 4mm 5mm 4mm, clip, 
width=0.5\textwidth]{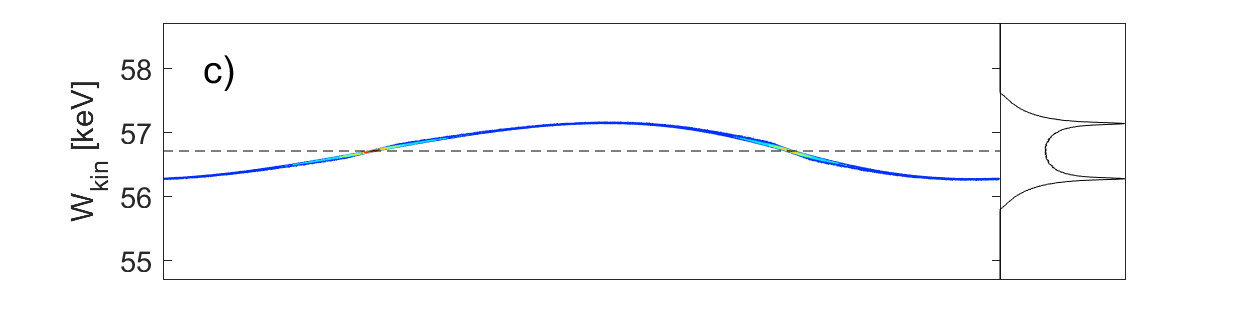}
\includegraphics[trim = 5mm 4mm 5mm 4mmm, clip, 
width=0.5\textwidth]{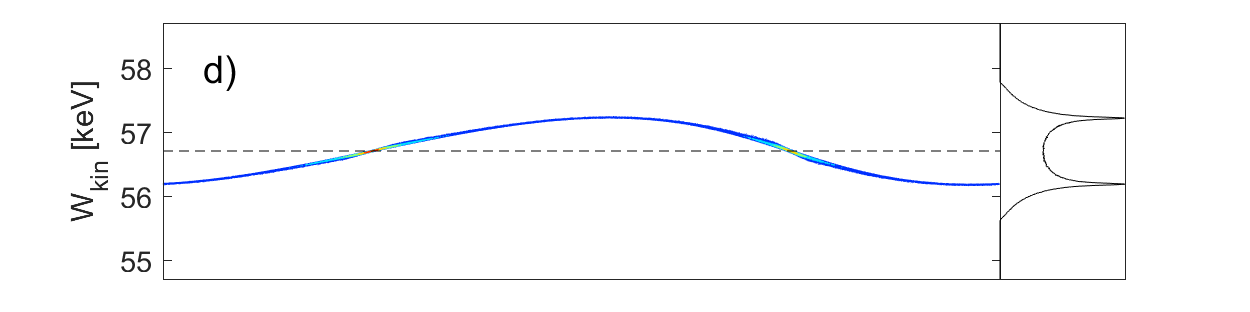}
\includegraphics[trim = 5mm 4mm 5mm 4mm, clip, 
width=0.5\textwidth]{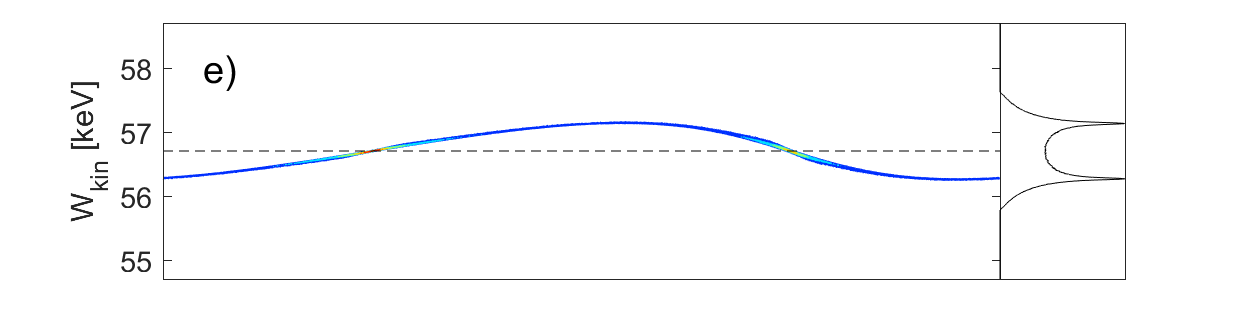}
\includegraphics[trim = 5mm 4mm 5mm 4mm, clip, 
width=0.5\textwidth]{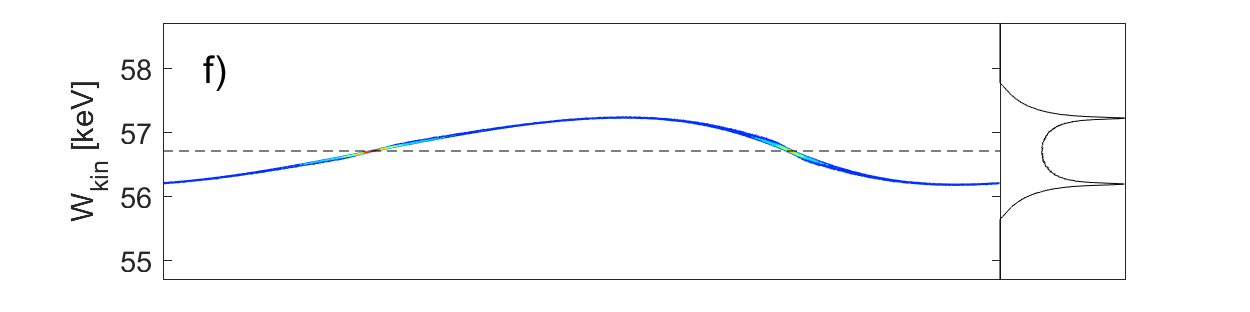}
\includegraphics[trim = 5mm 4mm 5mm 4mm, clip, 
width=0.5\textwidth]{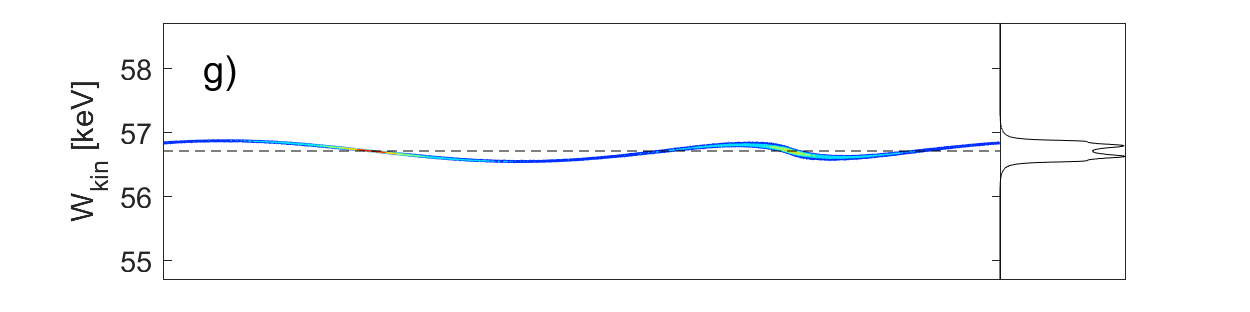}
\includegraphics[trim = 5mm 4mm 5mm 4mm, clip, 
width=0.5\textwidth]{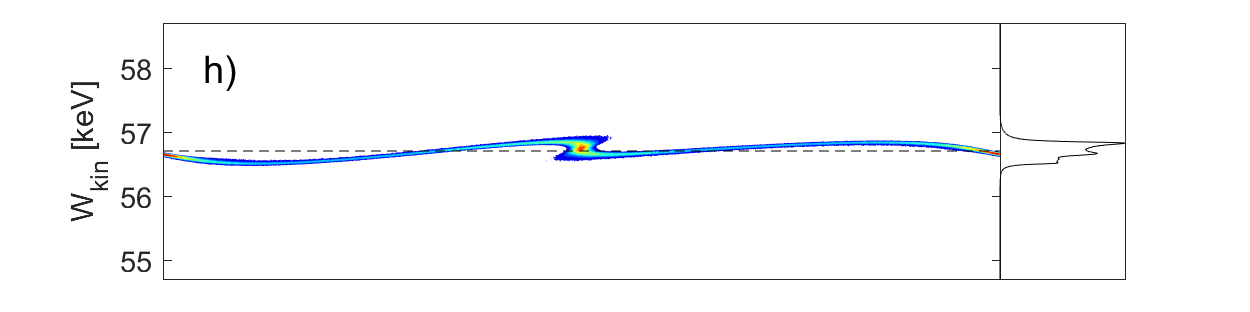}
\includegraphics[trim = 5mm 4mm 5mm 4mm, clip, 
width=0.5\textwidth]{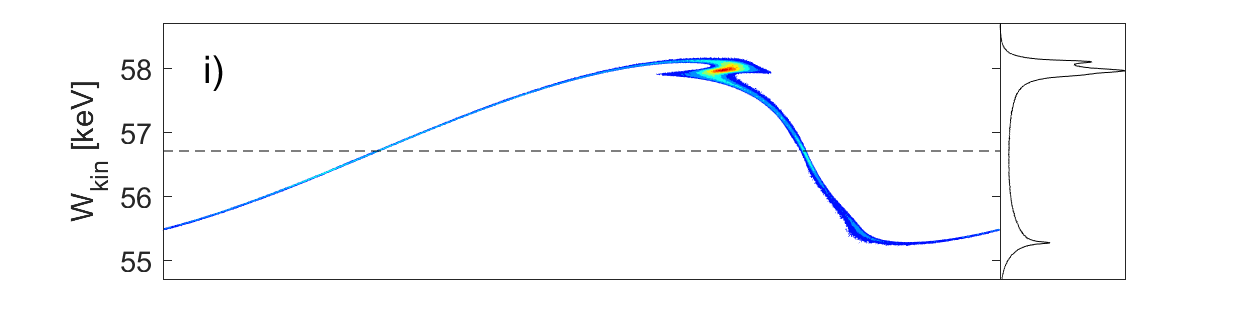}
\includegraphics[trim = 5mm 4mm 5mm 5mm, clip, 
width=0.5\textwidth]{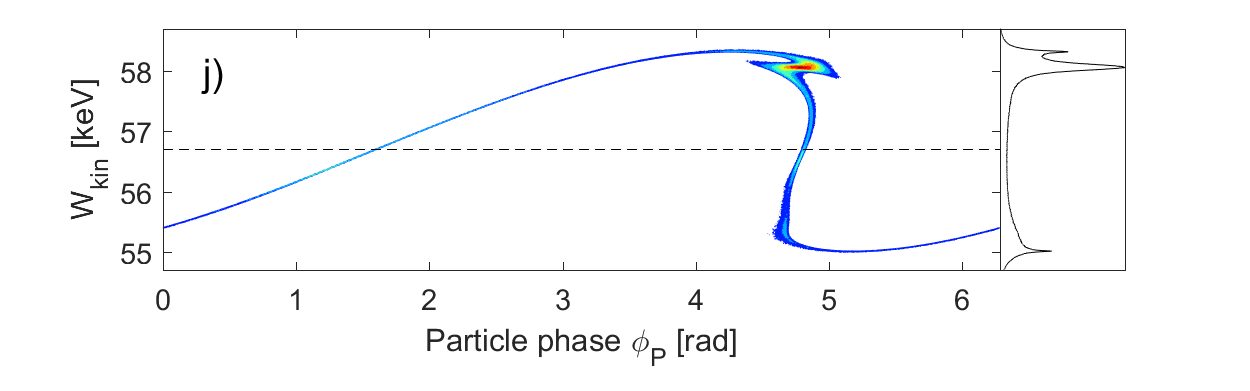}
\caption{Longitudinal phase space after each part of the structure a)-i) as indicated in Fig.~\ref{Picture}. Panel j) shows a slightly longer accelerator stage. The insets show the projected energy spectra in arb. linear units scaled to their maximum.}
\label{Fig:LongPhaseSpace}
\end{figure}
\begin{figure}[b]
\centering
\includegraphics[width=0.45\textwidth]{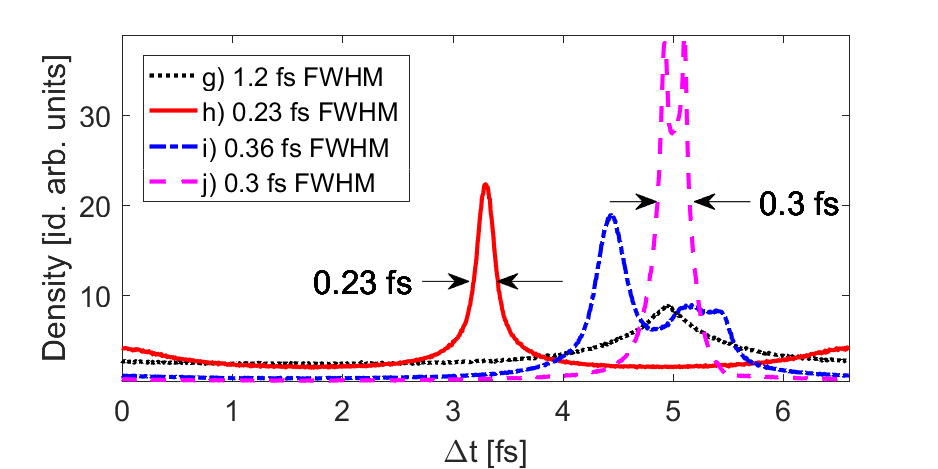}
\includegraphics[width=0.45\textwidth]{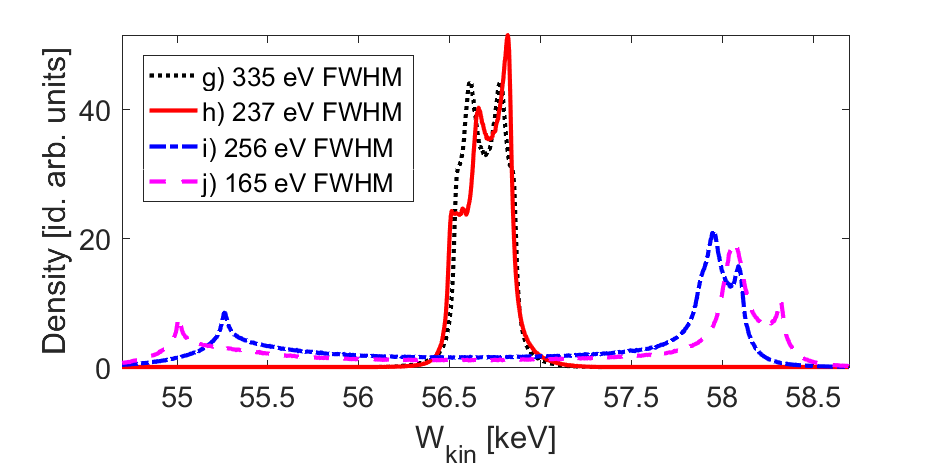}
\caption{Simulated bunch lengths (top panel) and energy spreads (bottom panel) along the structure. The arrival time difference $\Delta t=\phi_P/\omega$ is plotted for one laser period $T=\lambda/c=6.6$\,fs. The energy spectrum (h) is one cell into the second stage; the difference from (g) is mainly due to clipping particles at the edge of stage (2).}
\label{Bunchlength}
\end{figure}

Numerical simulations in DLAtrack6D~\cite{Niedermayer2017BeamScheme} are performed to describe the nonlinear dynamics of the experiment. The design principles can however be understood from simple analytic considerations. In each DLA cell, the energy gain is given by 
\begin{equation}
    \Delta W= q e_1 \lambda_g \cosh\left(\frac{\omega y}{\beta\gamma c}\right) \sin(\phi_P-\phi_S) 
    \label{Egain}
\end{equation}
where $\omega=2\pi c/\lambda$, $q=-e$ is the electron charge, $\phi_P=\omega\Delta t$ is the particle phase and $\phi_S$ is the synchronous phase. The laser field amplitude is characterized by the synchronous mode coefficient 
$e_1$, which we design to be 50~MV/m. Note that $qe_1$ is the on-crest gradient in the center of the channel. During $n$ DLA cells, the phase of an off-energy particle will slip as~\cite{Niedermayer2017BeamScheme}
\begin{equation}
    \Delta\phi_P = \frac{2\pi n}{\beta\gamma^3 m_ec^2}\Delta W.
    \label{PhiP}
\end{equation}
By means of a fractional period drift of $l_d$, the synchronous phase can be changed as
\begin{equation}
    \Delta \phi_S=2\pi\frac{l_d}{\lambda_g},
    \label{PhiS}
\end{equation}
while the change in the particle phase is negligible due to the shortness of this drift. We use $l_d=\lambda_g/2$ here, which leads to a $\pi$-shift of $\phi_S$; see the inset of Fig.~\ref{Picture} for the design of the synchronous phase and Fig.~\ref{Fig:LongPhaseSpace} for the longitudinal phase space evolution as computed by DLAtrack6D. The bunch profiles and energy spectra in identical arbitrary units are plotted in Fig.~\ref{Bunchlength}.
The combination of modulation and demodulation enables the device design as shown in Fig.~\ref{Picture}, which modulates the beam first, then transports it from (b) to (f), where alternating modulation/demodulation is employed for the purpose of transverse confinement. 
Then the segment (f)-(g) demodulates the beam, such that only a small residual energy spread remains. This remaining small energy spread finally compresses the already prebunched beam to a minimum bunch length roughly at the beginning of the second stage (h).
Simulation predicts that the bunch at point (h) is on the order of 230~as in length, with an energy spread of about 237~eV, i.e., this bunch would be suitable to inject into a scalable APF accelerator (cf. Eq.~\ref{Eq:Mathcing}). We only implement one segment of $28\,\mu$m length as the second stage here, which extends up to point (i). A slightly longer second stage would reduce the output energy spread to 165~eV at point (j), but this was not implemented in the experiment. 

The transverse focus of the bunched electrons is roughly in the center of stage 2. The $35~\mu$m drift between the stages is chosen as sufficiently large to spatially separate the laser beams (1) and (2) to avoid cross-talk. The phase difference between stage (1) and (2) can be chosen arbitrarily, thus the synchronous phase of stage (2) is arbitrary but held constant during measurement. The stage 2 synchronous phase in Fig.~\ref{Fig:LongPhaseSpace} was chosen such that the electron bunch drifts over the crest, producing a net energy gain of about 1.3~keV. Without this phase slippage, the energy gain would be $|qe_1|L=1.4$~keV.



The setup of the experiment is similar to~\cite{Black2019NetAccelerator}.
The DLA structures as seen in Fig.~\ref{Picture} are fabricated from 5$-$10 $\Omega$-cm B:Si. The elliptical pillars have dimensions of roughly $r_z=690$~nm, $r_y=830$~nm, and height $h=2.7~\mu$m, and the channel gap between the pillars is 300~nm.
The electron macro-bunch is produced by illumination of a silicon nanotip cathode~\cite{AndrewCeballos2019SiliconAccelerators}  with a laser of 1~$\mu$m wavelength and roughly 300~fs pulse length. The repetition rate is 100 kHz, which produces electron pulses at the same repetition rate and $730 \pm 100$~fs in length, measured by cross-correlation with the DLA drive laser. The electron beam is focused to a circular gaussian spot of width $230 \pm 30$ nm RMS at the beginning of the structure (a).

The outcoming electron beam is analyzed by a sector magnet spectrometer with a roughly 100~eV point spread function. The injection reference energy can be set in a range between 56~keV and 60~keV, with an energy spread less than 10 eV. Since the electron beam is operated with $\sim$ 300 electrons/sec, i.e., less than one electron per laser pulse, space charge is negligible. The energy spread is primarily limited by the power supply ripple of roughly 1 V at 60~kV. 

\begin{figure}[t]
\centering
\includegraphics[width=0.49\textwidth]{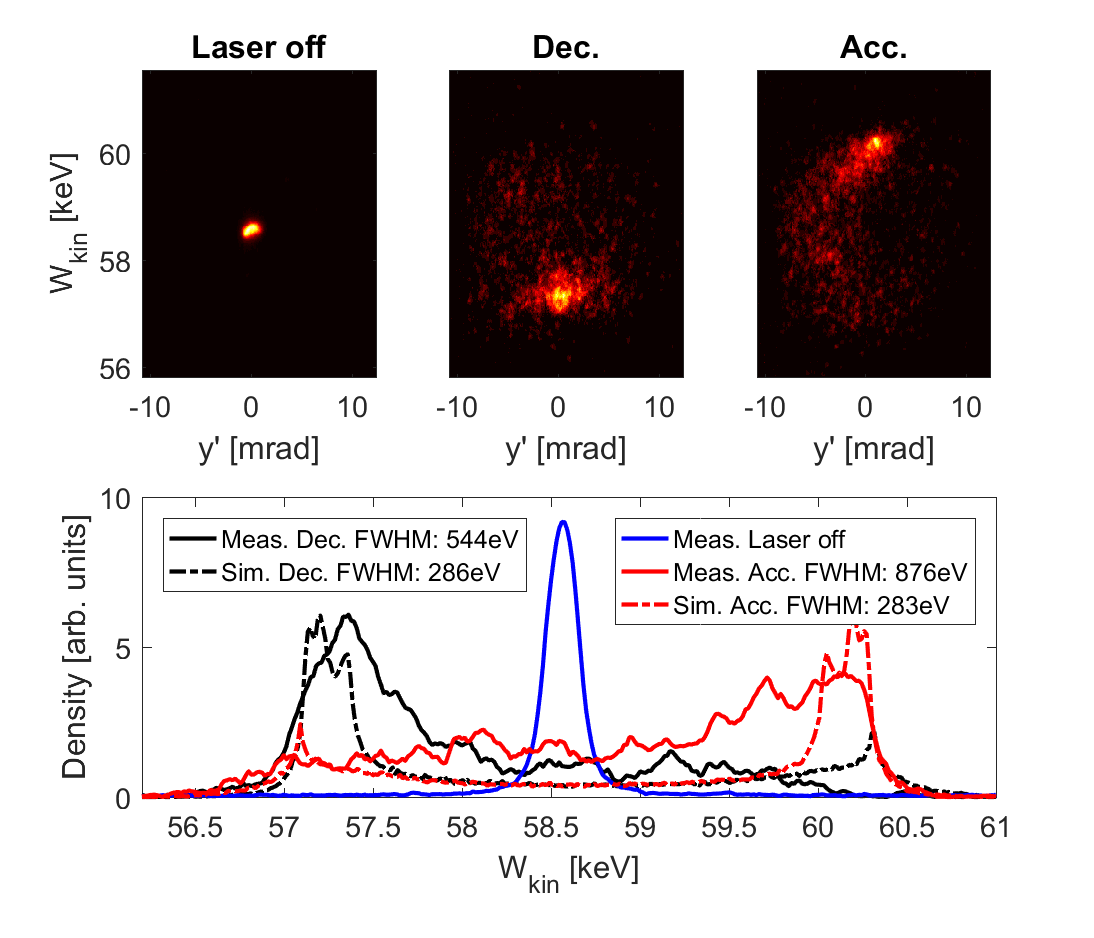}
\caption{Top panel: Data as recorded on the MCP screen for different stage 2 synchronous phases. The horizontal deflection axis is not significantly influenced by the spectrometer magnet. Bottom panel: Comparison of the respective spectra to simulation. We determine $1.5\pm0.1$~keV energy gain and $0.88_{-0.2}^{+0.0}$~keV FWHM energy spread at maximum gain.}
\label{Fig:HighEgain}
\end{figure}

The DLA structures are pumped with a commercial OPA system, driven by the same 1 $\mu$m regenerative amplifier that drives the cathode (also at 100~kHz). The pulse length is 605 $\pm$ 5 fs intensity FWHM at the chosen $\lambda=1980$~nm. The four DLA drive beams (see Fig.~\ref{Picture}) are focused to $1/e^2$~intensity radius of 22~$\pm$~1~$\mu$m. Each branch provides pulses ranging from 0 to $\sim$ 50~nJ, depending on the desired acceleration gradient. The phase of each branch is differentially controlled by (free-running, not locked) piezo delay stages. The phase stability is maintained to better than $\lambda/10$ over short timescales ($<1$ ~sec). The total averaging time per frame is limited by larger slow drifts.


\begin{figure}[t]
\centering
\includegraphics[width=0.5\textwidth]{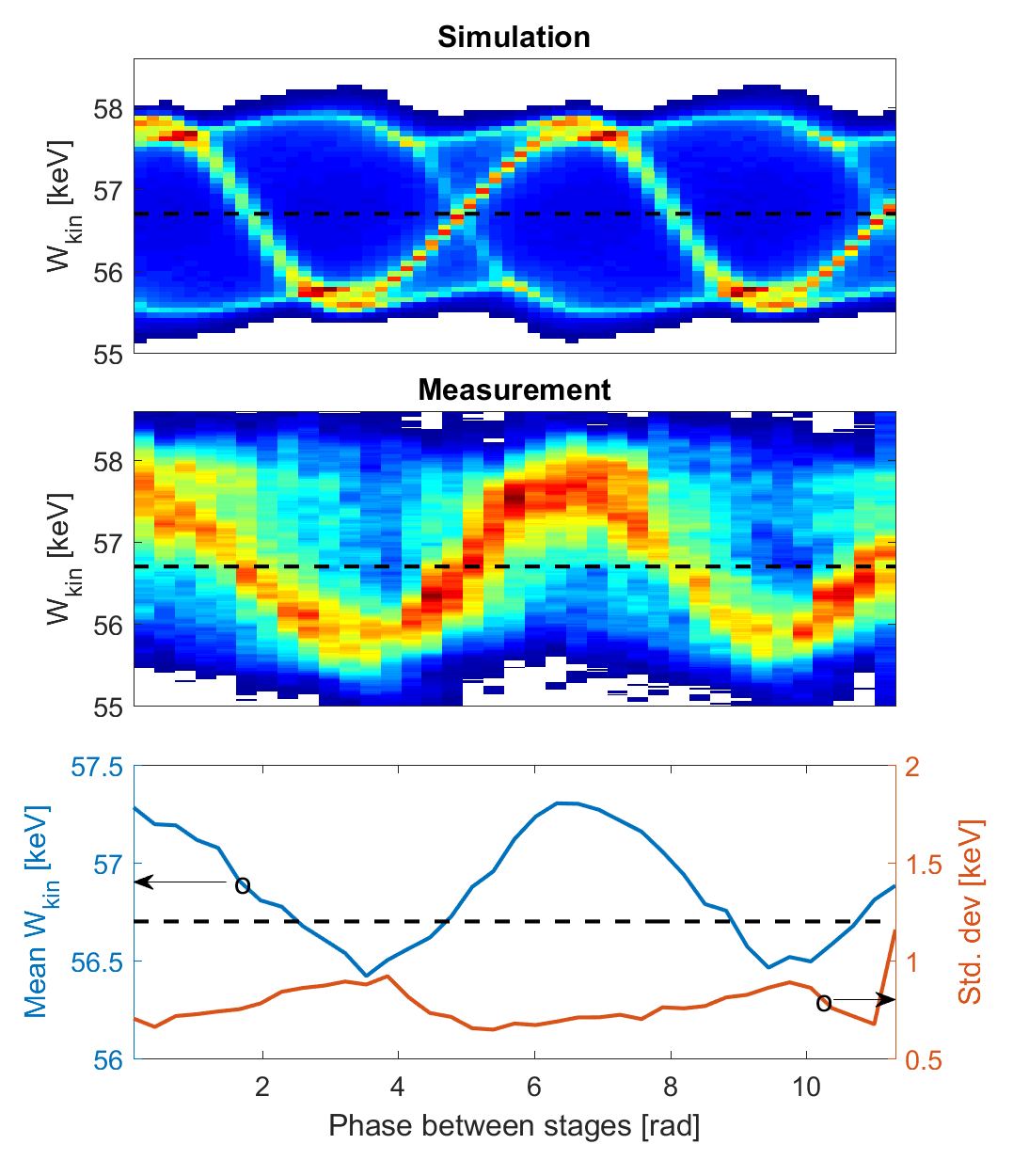}
\caption{The sinusoidal spectrogram shows excellent agreement with simulation, albeit with higher energy spread, which can be accounted to a vertical spread of $e_1(x)$ and the spectrometer point spread function. Bottom panel: Mean and standard deviation of the measured spectrogram.}
\label{Fig:BestDataSet}
\end{figure}

Both DLA stages are operated in the symmetric drive configuration, with zero relative phase between each stage's drive lasers (1a and 1b have the same phase, and 2a and 2b have the same phase). The injection phase of the electron into the second stage is controlled by symmetrically delaying the pairs of drive lasers (the phase of 2a and 2b are delayed keeping 1a and 1b fixed). Since accurate measurement of $e_1$ in stage 1 is not possible, we measure $e_1$ in stage~2 and assume the same power level at stage~1. Together with the phase error in stage~1 (laser 1a vs. 1b) this is the main source of driver errors.

Two example experimental spectra are shown in Fig.~\ref{Fig:HighEgain}, where the electron spectrum is coherently accelerated or decelerated depending on the injection phase into the second DLA stage. The increase of the transverse spot size is small, since the captured electrons spend roughly the same time on focusing and defocusing synchronous phases. The energy gain is $1.5\pm0.1$~keV, which indicates that the second stage was slightly overpowered. The energy spread is $0.88_{-0.2}^{+0.0}$~keV FWHM at maximum acceleration and $0.54_{-0.2}^{+0.0}$~keV FWHM at maximum deceleration. 

The main source of measurement error is the spectrometer point spread function, which shows a broader spectrum on the screen than reality. This is also visible in Fig.~\ref{Fig:HighEgain} in the laser-off curve exhibiting a width of about 0.2~keV FWHM. The energy spread is still larger than predicted by the simulations after accounting for the spectrometer point spread function, however. This excess energy spread is caused primarily by 3D geometry effects such as the finite height of the pillars and the mesa structure, leading to a possibly strong $e_1(x)$ dependency, see~\cite{Niedermayer2020ThreedimensionalAccelerators}. Consequently, there can be under-bunching, over-bunching, and vertical deflection in the same measurement. Additionally, there is non-uniformity in $e_1$ due to minor cross-talk between the drive lasers and intra-stage phase errors.

A full synchronous phase sweep measurement is shown in Fig.~\ref{Fig:BestDataSet} with $e_1=53\pm5$~MV/m, close to the design gradient. A clear sinusoidal spectral dependence is visible, i.e., the energy gain can be continuously selected by the inter-stage phase. Again, the source for the excess energy spread is the vertical spread of $e_1$ and due to the lower energy gain, there is significant background from non-trapped electrons. The centroid and standard deviation of the measurement data shown in the bottom panel indicate that the spread stays roughly constant at $~800\pm200$~eV, with slight increase at maximum deceleration, where some of the electrons were lost. Note that mean and std. dev. were taken over each entire spectrum including the decelerated tail. This causes the mean to be only about 0.6~keV above injection energy as compared to the 1.3~keV of the edge of the spectrum.

Together with the streaking experiment presented in~\cite{Black2019NetAccelerator}, we are now able to coherently move the electron beam in both dimensions (energy and deflection) on the microchannel plate (MCP) screen by changing the relative phase of the two stages. The observable small angle spread indicates that the beam leaves the structure without a large increase in divergence. For streaking it is advantageous to have a shorter second stage, which reduces phase slip errors. If only the energy spectrum is measured, a longer second stage (cf. Fig.~\ref{Fig:LongPhaseSpace} (j)) leads to higher energy gain and thus better relative resolution. Moreover, after slipping over the crest, the energy spread is additionally compressed at the expense of a slightly longer bunch length. The bunch length can be inferred from the measured energy spectra via comparison to simulations~\cite{Schonenberger2019GenerationAcceleration}, which we do not attempt here due to the large uncertainty for that measurement with our experimental parameters. 

In conclusion, we have demonstrated small energy spread bunching in a DLA based on a modulation-demodulation APF scheme. The resulting sub-femtosecond electron pulses were injected into a second stage for coherent acceleration. Good agreement with simulation results was achieved, with the increased experimental energy spread accounted for by three-dimensional field non-uniformity. This non-uniformity can be quite strong but can also be exploited to confine the beam vertically in future experiments, enabling a scalable APF DLA accelerator~\cite{Niedermayer2020ThreedimensionalAccelerators}. Preliminary full 3D particle tracking simulations indicate that APF buncher structures with vertical confinement would achieve similar energy spreads as predicted in the 2D simulations presented here. Due to the more involved longitudinal phase space manipulations as compared to a simple ballistic buncher, we call this device "fancy buncher".

\begin{acknowledgments}
This work is funded by the Gordon and Betty Moore Foundation (Grant GBMF4744).
\end{acknowledgments}

\bibliography{Mendeley}
\end{document}